\documentclass[pre, 10pt, aps, superscriptaddress, floatfix,
notitlepage, longbibliography, nofootinbib, nobibnotes]{revtex4-1}

\usepackage{amsmath}
\usepackage{amsfonts}
\usepackage{amssymb}
\usepackage{graphicx}
\usepackage{color}
\usepackage{xcolor}
\usepackage{sidecap}
\usepackage{float}
\sidecaptionvpos{figure}{b}
\usepackage[utf8]{inputenc}
\usepackage[mathscr]{euscript}
\usepackage{rotating}
\usepackage{comment}
\usepackage{wrapfig}
\usepackage{caption}
\usepackage{subcaption}
\usepackage{setspace}
\usepackage{bm}
\usepackage{tikz}
\usetikzlibrary{arrows.meta} 
\usetikzlibrary{calc}
\usepackage{tcolorbox}
\usepackage{booktabs}
\usepackage{lipsum}
\usepackage{hyperref}
\hypersetup{colorlinks=true,linkcolor=blue,citecolor=magenta,urlcolor=magenta}

\begin{document}
\title{Integrated Circular Polarisation Beam Shaper Based on Tilted Waveguide Bragg Gratings}

\author{Joel Warner}
\author{William J. Hughes}
\author{James C. Gates}
\author{Peter Horak}
\affiliation{Optoelectronics Research Centre, University of Southampton, Southampton, SO17 1BJ, United Kingdom}

\begin{abstract}
Scalable quantum platforms require optical systems that deliver well‑defined polarisation without relying on macroscopic bulk optics or manual alignment, motivating chip‑integrated approaches capable of shaping both the spatial mode and the polarisation of the emitted field. By developing a full three‑dimensional analytical model of the scattering process together with numerical simulations to design and evaluate the emitted field, we show that tilted waveguide Bragg gratings can act as integrated polarisation beam shapers. We find that scattering two co-propagating, orthogonally polarised pumps on a single grating is fundamentally limited: waveguide birefringence and the difference in scattering efficiency prevent full polarisation control. We overcome this with an orthogonal superimposed‑grating architecture in which two gratings driven independently by two pumps propagating perpendicular to each other succeed in generating any desired polarisation output apart from directions in a narrow spatial region. The approach provides a scalable, chip‑integrated route to polarisation‑engineered light delivery for quantum technologies.
\end{abstract}

\maketitle

\subsection*{Keywords}
Tilted Bragg gratings; Integrated photonics; Polarisation control; Waveguide beam shaping; Trapped-ion quantum computing; neutral-atom quantum computing; UV-written silica waveguides; Scalable quantum photonics; Out-of-plane grating couplers

% ============================================================
\section{Introduction}
% ============================================================

Trapped-ion and neutral-atom systems have emerged as leading platforms for quantum computation, simulation, and sensing, but scaling them beyond proof-of-principle demonstrations demands new methods of delivering optical fields to individual particles~\cite{blatt_entangled_2008}. Early ion traps used macroscopic blade or ring electrodes in which ions were addressed with freely propagating laser beams; more recent surface-electrode traps use electrode geometries fabricated on a planar chip, enabling lithographic control of the trapping potential and opening the possibility of co-integrating photonic elements on the same substrate~\cite{seidelin_microfabricated_2006, chiaverini_surface-electrode_2005, moehring_design_2011, allcock_microfabricated_2013, suleimen_surface_2024}. Similar developments have occurred in neutral-atom platforms, where atom chips have incorporated integrated optical functionality alongside on-chip magnetic and optical trapping structures~\cite{isichenko_photonic_2023, folman_ron_microscopic_2002, schmiedmayer_atom_2002, horak_possibility_2003}. In both cases, particles are held tens of microns to millimetres above the chip surface, typically 50--200\,\textmu m for surface-electrode ion traps, and extending to millimetre-scale distances in neutral-atom platforms including magneto-optical traps~\cite{niffenegger_integrated_2020, mehta_integrated_2016, mehta_integrated_2020, ballance_high-fidelity_2016}.

Delivering this light through bulk free-space optics remains the approach used in the majority of existing systems and imposes well-known constraints on alignment stability, optical access, and scalability~\cite{wineland_experimental_1998}. Integrated photonic devices offer a route to overcoming these limitations while also providing intrinsic phase stability as bulk beam paths accumulate vibration-induced phase noise, whereas on-chip paths share a common substrate and remain phase-locked. This makes phase‑sensitive control feasible in integrated platforms but more difficult in bulk systems. Pioneering demonstrations have shown that waveguides and grating couplers can be incorporated into ion-trap chips to route light directly to the trapping region~\cite{niffenegger_integrated_2020, mehta_integrated_2016, mehta_integrated_2020}. For many quantum applications, the delivered optical field must possess a well-defined polarisation state at the target location. Circular polarisation is particularly important because it selectively drives $\sigma^\pm$ transitions, enabling spin-state preparation, optical pumping into specific Zeeman sublevels, and qubit manipulations in trapped ions and neutral atoms~\cite{corwin_spin-polarized_1999, wineland_experimental_1998, corsetti_integrated-photonics-based_2026, tu_precision_2025, leu_polarization-insensitive_2024}. While linear polarisation can generally be produced straightforwardly using integrated beam-shaping structures, generating high-purity circular polarisation with the correct handedness above a chip surface remains challenging.

Tilted Bragg gratings (TBGs) written directly into waveguides offer a compact and lithographically defined route to free-space beam shaping. Through apodisation and phase engineering enabled by the UV-writing process, they provide control over both the radiation angle and tailored spatial profile of the generated beam~\cite{ahmed_direct_2021, ko_out--plane_2022, holmes_direct_2015}. While etched grating couplers have recently been explored for ion-trap integration~\cite{nshii_surface-patterned_2013, knollmann_collection_2026, lim_design_2021, krasnokutska_high_2019, shirao_high_2022, niffenegger_integrated_2020, kwon_multi-site_2024, kim_photonic_2018, lomonte_scalable_2024}, these devices have primarily focused on emission efficiency, beam shaping, and optical delivery to the trapping region, leaving the generation of high-purity circular polarisation largely unexplored. Likewise, the capability of TBGs to generate free-space beams with a well-defined polarisation state has not been systematically investigated.

The work presented here makes two contributions that expand the capabilities of integrated photonic beam-shaping with TBGs. First, we develop the full 3D scattering formalism for tilted Bragg gratings with arbitrary in‑plane rotation, revealing polarisation‑mixing mechanisms that are absent from existing 2D treatments. Second, we use this framework to analyse a single-TBG beam shaper, identifying the constraints that limit its ability to preserve circular polarisation, and introduce a novel superimposed orthogonal‑grating structure that enables robust generation of circularly polarised free‑space beams directly from a planar silica waveguide. Together, these results establish a physically complete description of TBG emission and demonstrate a practical route toward scalable, polarisation‑engineered light delivery for quantum technologies. Section~\ref{sec:theory} presents the device geometry and scattering model. Section~\ref{sec:design} describes the grating design procedure, pump depletion, and polarisation metric. Section~\ref{sec:single} analyses the single-TBG architecture.
Section~\ref{sec:ortho} presents the orthogonal architecture and analyses its parasitic scattering channels.

\section*{Method}

% ============================================================
\section{Device Geometry and 3D Scattering Model}
\label{sec:theory}
% ============================================================

\subsection{Coordinate System and Waveguide}

\begin{figure}[h]    
    \centering

    %-----------------------------------------
    % Left: big 3D schematic
    %-----------------------------------------
    \begin{subfigure}[t]{0.55\textwidth}
    \centering
    \begin{minipage}[c][6cm][c]{\linewidth}
        \centering
        \includegraphics[width=\linewidth]{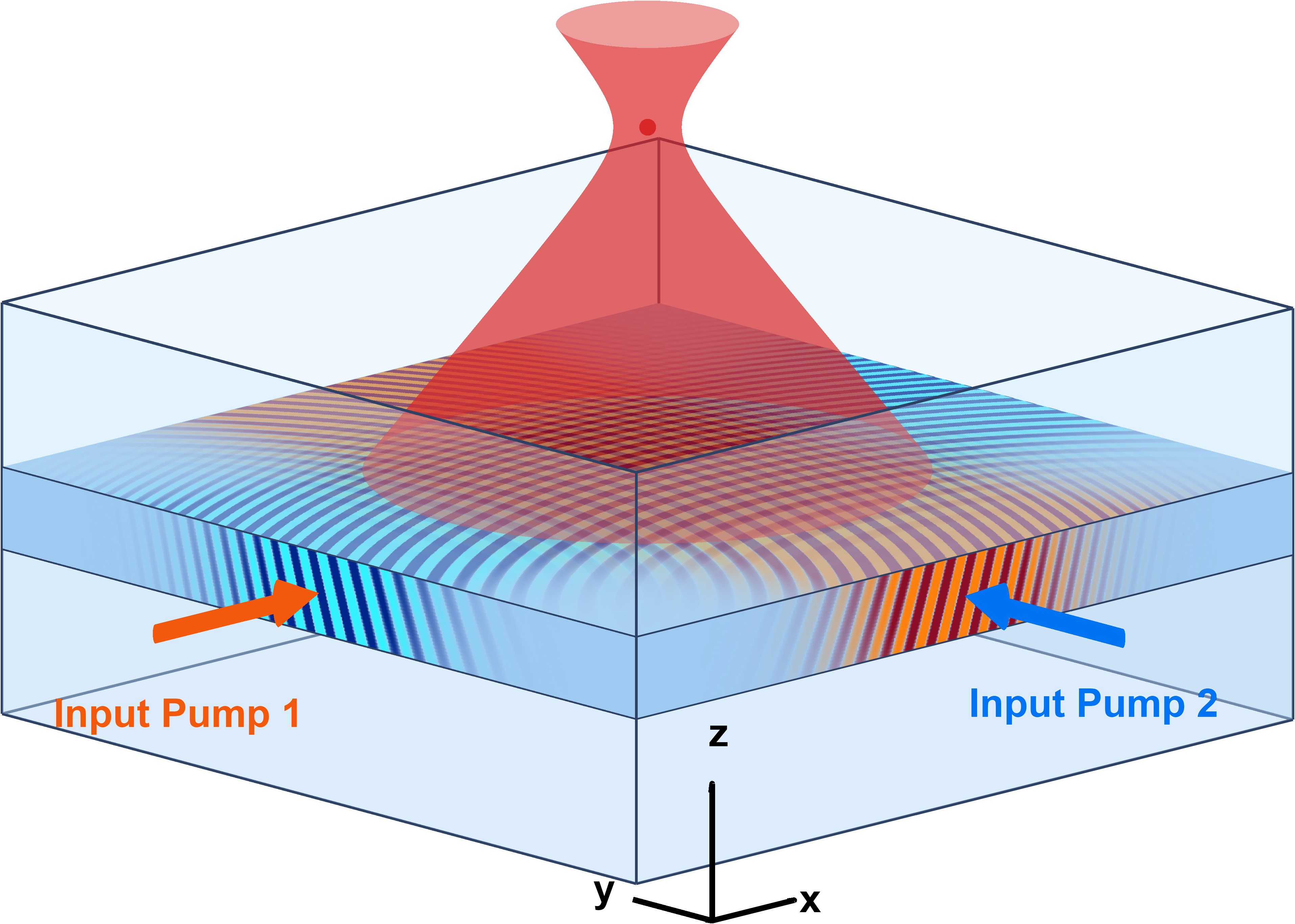}
    \end{minipage}
    \vspace{1.5em}
    \caption{3D schematic of tilted Bragg grating}
    \label{fig:3d_schematic}
    \end{subfigure}
    \hfill
    %-----------------------------------------
    % Right column: side view + top view stacked
    %-----------------------------------------
    \begin{minipage}[t]{0.40\textwidth}

        %--- Side view ---
        \vspace{-3.25cm}
        \begin{subfigure}[t]{\textwidth}
            \centering
            \begin{tikzpicture}[scale=1.6]
                \definecolor{chipblue}{RGB}{185,216,244}
                \definecolor{cladblue}{RGB}{210,225,255}
            
                %----------------------------------------
                % Vertical stack (centred core)
                %----------------------------------------
                % Lower cladding:  -1.55 → -1.30
                \fill[cladblue] (-2,-1.55) rectangle (2,-1.30);
                \draw[line width=0.12pt, cladblue!60!black] (-2,-1.55) -- (2,-1.55);
            
                % Core (chip):     -1.30 → -0.30
                \fill[chipblue] (-2,-1.30) rectangle (2,-0.30);
                \draw[line width=0.16pt] (-2,-1.30) -- (2,-1.30);
                \draw[line width=0.16pt] (-2,-0.30) -- (2,-0.30);
                \node at (1.65,-0.45) {chip};
            
                % Upper cladding:  -0.30 → -0.05
                \fill[cladblue] (-2,-0.30) rectangle (2,-0.05);
                \draw[line width=0.12pt, cladblue!60!black] (-2,-0.05) -- (2,-0.05);
                \node at (1.45,-0.18) {cladding};
            
                % Left slice
                \draw[line width=1.2pt, blue!60] (-1.25,-1.30) -- (-0.35,-0.30);
            
                % Centre slice
                \draw[line width=1.2pt, blue!60] (-0.35,-1.30) -- (0.55,-0.30);
            
                % Right slice
                \draw[line width=1.2pt, blue!60] (0.50,-1.30) -- (1.40,-0.30);
            
                %----------------------------------------
                % Tilt angle reference
                %----------------------------------------
                \draw[dashed] (0.05,-1.30) -- (0.05,-0.30);
            
                % Angle arc
                \draw[->] (0.05,-0.35) arc (90:25:0.30);
                \node[right] at (0.02,-0.55) {$\theta$};
            
                %----------------------------------------
                % Grating period label
                %----------------------------------------
                \draw[<->] (-1.15,-1.17) -- (-0.25,-1.17);
                \node[below] at (-0.68,-0.9) {$\Lambda$};
            
                %----------------------------------------
                % Guided mode arrow
                %----------------------------------------
                \draw[->,very thick,red] (-2.0,-0.80) -- (-1.2,-0.80);
                \node[left, red] at (-1.22,-0.45) {Input};
                \node[left, red] at (-1.05,-0.65) {Pump 1};
            
                \begin{scope}[shift={(1.4,-1.1)}, scale=0.55]
                    \draw[->, line width=0.4pt] (0,0) -- (0.6,0) node[right] {$x$};
                    \draw[->, line width=0.4pt] (0,0) -- (0,0.6) node[above] {$z$};
                    \node at (0,0) {\Large$\otimes$}; % y into page
                    \node[right] at (-0.6,0.0) {$y$};
                \end{scope}
            
            \end{tikzpicture}

            \caption{Side view Grating A}
            \label{fig:side_view}
        \end{subfigure}

        \vspace{0.15cm}

        %--- Top view ---
        \begin{subfigure}[t]{\textwidth}
            \centering
            \begin{tikzpicture}[scale=1.6]
                \definecolor{chipblue}{RGB}{219,234,248};
                \fill[chipblue] (-2,-1) rectangle (2,1);
                \draw[line width=0.16pt] (-2,-1) -- (2,-1);
                \draw[line width=0.16pt] (-2,1) -- (2,1);
                \draw[line width=0.16pt] (-2,-1) -- (-2,1);
                \draw[line width=0.16pt] (2,-1) -- (2,1);
            
                \node at (1.65,0.8) {chip};
            
                %--- 3D offset amount ---
                \def\dx{-0.15}
                \def\dy{0.35}
            
                %--- Move slice downward ---
                \def\shift{-0.15}
            
                %--- Front edge of grating slice ---
                \coordinate (A) at (-1.5, -0.5 + \shift);
                \coordinate (B) at ( 1.5,  0.5 + \shift);
            
                %--- Back edge ---
                \coordinate (A2) at ($ (A) + (\dx,\dy) $);
                \coordinate (B2) at ($ (B) + (\dx,\dy) $);
            
                %--- Shadow offset ---
                \def\sx{-0.02}
                \def\sy{-0.09}
            
                %--- Shadow coordinates ---
                \coordinate (As) at ($ (A) + (\sx,\sy) $);
                \coordinate (Bs) at ($ (B) + (\sx,\sy) $);
                \coordinate (A2s) at ($ (A2) + (\sx,\sy) $);
                \coordinate (B2s) at ($ (B2) + (\sx,\sy) $);
            
                %--- Shadow behind slice ---
                \fill[black!40, opacity=0.25] (As) -- (Bs) -- (B2s) -- (A2s) -- cycle;
            
                %--- Actual slice ---
                \fill[blue!20] (A) -- (B) -- (B2) -- (A2) -- cycle;
                \draw[blue!60, line width=0.8pt] (A) -- (B);
                \draw[blue!60, line width=0.8pt] (A2) -- (B2);
                \draw[blue!60, line width=0.6pt] (A) -- (A2);
                \draw[blue!60, line width=0.6pt] (B) -- (B2);
            
                %--- Reference axis ---
                \draw[dashed] (-0.4,0.0) -- (1.0,0.0);
                \node at (0,0) {$\odot$};
                %--- measurement line ---
                \draw[dashed] (0,0) -- (0.95,0.33);
            
                %--- Angle label ---
                \draw[->] (0.85,0.0) arc (-30:30:0.3);
                \node[right] at (1.0,0.15) {$\phi$};
            
                %--- Label ---
                \node[above,blue!60] at (-1.25,-0.15 + 0.2) {grating slice};
            
                %--- Guided mode ---
                \draw[->,very thick, red] (0,-1.3) -- (0,-0.55);
                \node[below, red] at (0,-1.3) {Input Pump 1};

                % Axis triad using dot/cross notation
                \begin{scope}[shift={(1.6,-0.8)}, scale=0.55]
                    \draw[->, line width=0.4pt] (0,0) -- (-0.6,0) node[left] {$y$};
                    \draw[->, line width=0.4pt] (0,0) -- (0,0.6) node[above] {$x$};
                    \node at (0,0) {\Large$\odot$}; % z out of page
                    \node[right] at (0.2,0.0) {$z$};
                \end{scope}

            \end{tikzpicture}

            \caption{Top view Grating A}
            \label{fig:top_view}
        \end{subfigure}

    \end{minipage}

    \caption{Geometry of the tilted waveguide Bragg grating.
    (a)~Three-dimensional schematic showing the single-grating configuration
    (grating A, Pump 1, Orange) and the superimposed orthogonal architecture obtained by adding
    grating B (Pump 2, Blue). The guided pumps are scattered by the
    TBGs towards a common target above the chip.
    (b)~Side view defining the grating tilt $\theta$ and period $\Lambda$ (shown for the case $\phi=0$).
    (c)~Top view defining the in-plane rotation $\phi$.}
\label{fig:schematic_figure}
\end{figure}

Fig.~\ref{fig:schematic_figure} shows the device geometry of the tilted waveguide Bragg grating (TBG) beam shaper considered throughout this work. The device is made of a photosensitive doped silica layer, into which a periodic refractive-index modulation has been inscribed by UV exposure. This layer is fabricated between two pure silica cladding layers. 
We define $x$ as the waveguide propagation axis for pump A, $y$ as the in-plane transverse axis, and $z$ as the vertical, out-of-plane axis. The waveguide supports two guided modes relevant to this work: the TE mode, whose dominant electric field component lies along $y$ for grating A, and the TM mode, whose dominant electric field component lies along $z$. There is a small birefringence between these input modes, characterised by the effective index difference $\Delta n_\text{eff} = n_\text{eff}^\text{TE} - n_\text{eff}^\text{TM}$. For pump B, used in Sec.\ \ref{sec:ortho}, the roles of $x$ and $y$ are exchanged.

The UV-inscribed index modulation forms the physical tilted Bragg
grating, whose geometry is characterised by two independent angular
degrees of freedom, shown in Fig.~\ref{fig:schematic_figure}(b) and~(c).
The first is the out-of-plane tilt angle $\theta$, which sets the
inclination of the grating planes away from the vertical $z$ axis. The second degree of
freedom is the in-plane rotation angle $\phi$, defined in the $xy$ plane
as the angle between the grating slice and the $y$ axis, shown in Fig.~\ref{fig:schematic_figure}(c). Together, $\theta$ tilts
the grating planes out of the waveguide plane while $\phi$ rotates them
about the vertical axis, so that a given plane's orientation is fixed by
both angles; the side view of Fig.~\ref{fig:schematic_figure}(b) shows
the tilt $\theta$ for the $\phi = 0$ case. The grating period $\Lambda$,
also indicated in Fig.~\ref{fig:schematic_figure}(b), is the spacing of
the grating planes along the in-plane grating vector.

%%%%%%%%%%%%%%%%%%%%%%%%%%%%%%%%%%%%%%%%%%%%%%%

\subsection{Scattering Coefficient}
\label{sec:alpha_s}

The scattering coefficient $\alpha_s$, defined as the fraction of guided
power diffracted from the waveguide per unit length, determines the local
conversion of the pump into the free-space field. For the 2D geometry,
where the pump propagates along $x$ and the grating is tilted only in the
$xz$ plane ($\phi=0$), this coefficient was derived by
Yoshino~\cite{yoshino_theoretical_2012} for $\theta=45^\circ$ and extended
to arbitrary tilt by Ko et al.~\cite{ko_out--plane_2022}. Using the same approach but extended to 3D gives
\begin{equation}
    \alpha_{s} =
        \underbrace{
        \frac{\sqrt{2\pi}}{w_0}
        \left( \frac{1}{2\cos^2\theta_{\mathrm{inc}}} \right)^{2}
        \left( \frac{\Delta n_g}{n_{\mathrm{eff}}} \right)^{2}
        }_{\text{Fresnel reflectivity}}
        \;
        \underbrace{\left( \frac{\pi}{\Lambda} \right)^{2}
        \left( \frac{w_0}{\sin 2\theta} \right)^{2}
        }_{\text{interaction length}}
        \;
        \underbrace{
        \exp\!\left[
        -\left( \frac{w_0}{\sin 2\theta} \right)^{2}
        \left( \beta'\cos^{2}\theta - \frac{\pi}{\Lambda} \right)^{2}
        \right]
        }_{\text{Bragg resonance}}
        \;
        \sin\gamma,
    \label{eq:alpha_s}
\end{equation}
where $w_0$ is the Gaussian waveguide-mode waist, $\Lambda$ is the
grating period, $\Delta n_g$ is the refractive-index modulation depth,
$n_{\mathrm{eff}}$ is the guided-mode effective index, and $\gamma$ is
the angle between the scattered beam and the waveguide axis. The grating
geometry enters through
\begin{equation}
\hat{\mathbf{n}}_g =
\begin{pmatrix} -\cos\theta\cos\phi \\ \sin\theta \\ -\cos\theta\sin\phi \end{pmatrix},
\quad
\cos\theta_\mathrm{inc} = |\hat{\mathbf{k}}_\mathrm{pump}\cdot\hat{\mathbf{n}}_g|,
\quad
\beta' = \mathbf{k}_\mathrm{pump}\cdot\hat{\mathbf{n}}_g,
\quad
\mathbf{k}_\mathrm{pump} = \beta\,\hat{\mathbf{k}}_\mathrm{pump} = (\beta,0,0).
    \label{eq:geometry}
\end{equation}
The 3D scattering coefficient, Eq.~\eqref{eq:alpha_s}, now contains two distinct angles: the geometric tilt $\theta$ of each grating plane, as illustrated in Fig.~\ref{fig:schematic_figure}(b), and the incidence angle $\theta_\mathrm{inc}$, which represents the angle at which the pump hits the grating planes. In the 2D geometry the pump propagates along $x$ and meets each plane at exactly the tilt angle, so $\theta_\mathrm{inc} = \theta$. Once the grating is rotated in-plane ($\phi\neq 0$), the same pump wavevector intersects the grating planes at a different angle, such that $\theta_\mathrm{inc}$ departs from $\theta$ according to Eq.~\eqref{eq:geometry}.

The three bracketed groups in Eq.~\eqref{eq:alpha_s} have a clear
physical interpretation.
%########
The Fresnel factor $(2\cos^2\theta_\mathrm{inc})^{-2}$ represents the single-plane reflectivity for $s$-polarised light, derived from the standard Fresnel coefficient to first order in the index contrast ($\Delta n_g / n_\mathrm{eff} \ll 1$). The use of $\theta_\mathrm{inc}$ accounts for the local pump angle when $\phi \neq 0$. The interaction length factor $\left( \frac{\pi}{\Lambda} \right)^{2} (w_0/\sin 2\theta)^2$ quantifies the number of planes intersected before the ray exits the waveguide, which is minimised at $\theta = 45^\circ$. Finally, the Bragg resonance factor (the Gaussian exponential in Eq.~\eqref{eq:alpha_s}) peaks when the grating period and the orientation of each grating plane direct the reflected ray into the same Bragg-selected propagation direction. Under this condition, the local reflection from each plane is maximised because the plane tilt reinforces the direction set by the grating period.
%########

For $p$-polarised incidence, the Fresnel coefficient introduces the
additional angular factor
\begin{equation}
    \alpha_p
    = \alpha_s \cos^2(2\theta_\mathrm{inc}),
    \label{eq:alpha_p}
\end{equation}
so that $\alpha_p<\alpha_s$ for
$\theta_\mathrm{inc}\neq0$ and vanishes as
$\theta_\mathrm{inc}\rightarrow45^\circ$. This Brewster-like suppression
restricts the scattering efficiency of the $p$ channel and becomes
important for the single-grating architecture of
Section~\ref{sec:single}.

In-plane rotation also changes the polarisation basis of the scattering
process. The local $s$ (perpendicular to scattering plane) and $p$ (in the scattering plane) directions are defined by
\begin{align}
    \hat{\mathbf{e}}_s &=
    \frac{\hat{\mathbf{n}}_g\times\hat{\mathbf{k}}_\mathrm{pump}}
    {|\hat{\mathbf{n}}_g\times\hat{\mathbf{k}}_\mathrm{pump}|},
    \label{eq:es} \\
    \hat{\mathbf{e}}_p &=
    \frac{\hat{\mathbf{e}}_s\times\hat{\mathbf{k}}_\mathrm{pump}}
    {| \hat{\mathbf{e}}_s\times\hat{\mathbf{k}}_\mathrm{pump} |},
    \label{eq:ep}
\end{align}
so a TE pump,
$\mathbf{E}_\mathrm{pump}=E_0\hat{\mathbf{y}}$, has components
\begin{equation}
    E_s = E_0(\hat{\mathbf{y}}\cdot\hat{\mathbf{e}}_s),
    \qquad
    E_p = E_0(\hat{\mathbf{y}}\cdot\hat{\mathbf{e}}_p).
\end{equation}
At $\phi=0$, $\hat{\mathbf{e}}_s=\hat{\mathbf{y}}$ and the TE pump couples
purely through the $s$ channel. For $\phi\neq0$, the scattering plane
rotates relative to the fixed laboratory-frame pump polarisation and both
components become non-zero. The total local scattering rate is therefore
\begin{equation}
    \alpha_\mathrm{total}
    =
    \left(\hat{\mathbf{y}}\cdot\hat{\mathbf{e}}_s\right)^2\alpha_s
    +
    \left(\hat{\mathbf{y}}\cdot\hat{\mathbf{e}}_p\right)^2\alpha_p,
    \label{eq:alpha_total}
\end{equation}
with $\alpha_s$ and $\alpha_p$ evaluated at the local
$\theta_\mathrm{inc}$ and $\beta'$.

\begin{figure}[t]
    \centering
    \includegraphics[width=\linewidth]{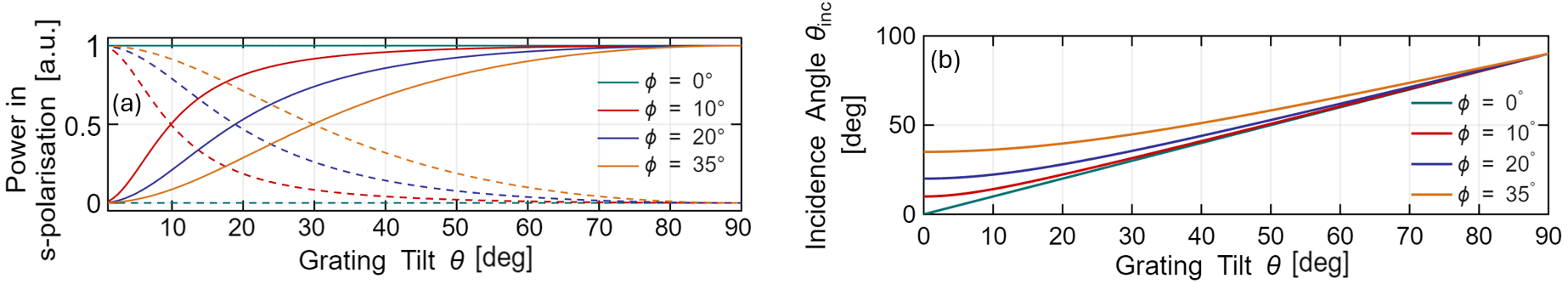}
    \caption{(a) Fraction of input pump power coupling into the $s$
    channel, $(\hat{\mathbf{y}}\cdot\hat{\mathbf{e}}_s)^2$, as a
    function of grating tilt $\theta$ for several values of in-plane
    rotation $\phi$. Solid lines represent a TE ($y$-polarised) input
    pump; dashed lines a TM ($xz$-plane) input pump. At $\phi = 0$ the
    TE pump couples purely into the $s$ channel for all tilt angles,
    recovering the 2D result. Increasing $\phi$ introduces mixing
    between the $s$ and $p$ channels with a fraction that depends on
    both $\phi$ and $\theta$. (b) Incidence angle $\theta_\mathrm{inc}$ as a function of grating tilt $\theta$ for the same in-plane rotations.}
    \label{fig:sp_mixing}
\end{figure}

To quantify the mixing, we examine the fraction of TE pump power in the $s$ channel, $(\hat{\mathbf{y}} \cdot \hat{\mathbf{e}}_s)^2$, in Fig.~\ref{fig:sp_mixing}(a) and the incidence angle $\theta_\mathrm{inc}$ in Fig.~\ref{fig:sp_mixing}(b) as functions of grating tilt $\theta$ for fixed values of $\phi$. At $\phi = 0$, the 2D limit is recovered: the TE pump couples entirely into the $s$ channel, and $\theta_\mathrm{inc} = \theta$. As $\phi$ increases, the scattering plane rotates relative to the
laboratory-frame TE polarisation, distributing the input power between
the local $s$ and $p$ channels. Since $\alpha_p < \alpha_s$, this transfer also reduces scattering efficiency, while the concurrent increase in $\theta_\mathrm{inc}$ above $\theta$ further weakens $p$-channel coupling via the Brewster factor. Consequently, precise polarisation control is challenging because the non-zero $\phi$ required for focusing light across the grating aperture causes local scattering axes to rotate with the geometry, resulting in angle-dependent projections onto both channels even for a nominally pure TE input.

% ============================================================
% ============================================================
\section{Grating Design and Polarisation}
\label{sec:design}
% ============================================================

\subsection{Grating Design}
\label{sec:grating_design}

Having established how the local scattering strength depends on the grating
geometry through the coefficient $\alpha_s$, we now describe how the
geometry and modulation profile are chosen to produce an arbitrary target
field above the chip. Throughout this work we take the target to be a
focused Gaussian beam, but the method works for any target field in the
observation plane. The grating is designed to produce the target from the
TE pump alone. Circular polarisation is then produced by launching a
second, TM, pump through the same grating with a chosen relative amplitude
and phase, as developed in Section~\ref{sec:single}.

The polarisation of each scattered channel is fixed by the pump
polarisation and the output direction through the coupling of
Section~\ref{sec:alpha_s}, and is therefore a known quantity rather than a
free one. The design does not shape polarisation directly: it matches the
scalar target amplitude for the TE channel, and the polarisation delivered
to the target is set afterwards by combining this channel with the
independently launched TM pump. The fields in the design below are treated
as scalar accordingly.

The design begins from the desired output field profile. A target field
$E_\mathrm{tar}$ is defined in the observation plane and propagated back
to the grating plane $z = 0$ using the angular spectrum method, giving the complex field
$E_\mathrm{grat}(x,y)$ that the grating must reproduce at each point. Its
local phase gradient fixes the in-plane components of the required output
wavevector,
\begin{equation}
    k_x = \frac{\partial \arg(E_\mathrm{grat})}{\partial x}, \qquad
    k_y = \frac{\partial \arg(E_\mathrm{grat})}{\partial y},
    \label{eq:kxky}
\end{equation}
and the out-of-plane component follows from the requirement that the
scattered light be a propagating wave,
\begin{equation}
    k_z = \sqrt{n_{0}^2k_0^2 - k_x^2 - k_y^2}.
    \label{eq:kz}
\end{equation}
The full output wavevector $\mathbf{k}_\mathrm{out}=(k_x,k_y,k_z)$ and therefore the emission direction is set by the phase of the back‑propagated target field. The required grating geometry is obtained from the vector Bragg condition. The grating vector is $\mathbf{K}_\mathrm{grat}=\mathbf{k}_\mathrm{in}-\mathbf{k}_\mathrm{out}$ with $\mathbf{k}_\mathrm{in}=(\beta,0,0)$. Its in‑plane components determine the local grating period and rotation, $\Lambda(x,y)=2\pi/\sqrt{K_x^2+K_y^2}$ and $\phi(x,y)=\arg(K_x+iK_y)$, while the tilt $\theta(x,y)$ is chosen to maximise the scattering efficiency Eq.~\eqref{eq:alpha_s}. In general, all three parameters vary across the aperture to steer and focus the beam.

With the geometry fixed, we next determine the spatial shape of the
refractive-index modulation required to reproduce the target amplitude.
The scattered field produced for unit index modulation,
$E_\mathrm{unit}(x,y)$, is evaluated from the scattering model of
Section~\ref{sec:alpha_s}. Comparing this with the back-propagated target
field $E_\mathrm{grat}(x,y)$ gives the unnormalised modulation requirement.
We define the normalised zero-depletion modulation shape as
\begin{equation}
    g(x,y)
    =
    \frac{
    |E_\mathrm{grat}(x,y)|/|E_\mathrm{unit}(x,y)|
    }{
    \displaystyle
    \max_{x,y}
    \left[
    |E_\mathrm{grat}(x,y)|/|E_\mathrm{unit}(x,y)|
    \right]
    },
    \qquad
    \max_{x,y} g(x,y)=1.
    \label{eq:dng_shape}
\end{equation}
Because the target is scalar while the scattered field carries
polarisation, the field modulus is matched here. Equation~\eqref{eq:dng_shape}
therefore specifies only the spatial shape of the grating modulation; its
absolute magnitude is set by the required pump depletion, as described
below.

\subsection{Pump Depletion}
\label{sec:depletion}

The modulation profile $g(x,y)$ defined in Eq.~\eqref{eq:dng_shape} specifies the zero‑depletion grating strength; we now determine how this profile must be scaled when the pump depletes along the grating. As the guided pump propagates through the grating, power is transferred to the scattered field and the remaining pump amplitude decreases. The modulation must therefore increase along the propagation direction to
maintain the required scattered-field envelope.

We write the normalised zero-depletion modulation profile obtained from
Eq.~\eqref{eq:dng_shape} as $g(x,y)$, with $\max(g)=1$. The grating is designed for the TE pump. We therefore denote by
$\tilde{\alpha}_\mathrm{TE}(x,y)$ the grating's local scattering coefficient of the TE pump evaluated for unit index modulation ($\Delta n_g=1$). The physical local scattering coefficient is therefore $\Delta n_g^2\tilde{\alpha}_\mathrm{TE}$. The cumulative scattering demand along $x$ at transverse position $y$ is
\begin{equation}
    S(x,y)
    =
    \int_0^x
    g(x',y)^2
    \tilde{\alpha}_\mathrm{TE}(x',y)
    \,\mathrm{d}x' .
    \label{eq:cumulative}
\end{equation}
Defining $S_\mathrm{max}=\max_y S(L_\mathrm{grat},y)$, the modulation scale
for complete depletion of the most strongly scattering line $y$ is
$\bar n_\mathrm{max}=1/\sqrt{S_\mathrm{max}}$. A desired depletion fraction
$D$ is then imposed through
$\bar n=\bar n_\mathrm{max}\sqrt{D}$.

For an input pump distribution $P_0(y)$, normalised to unit peak power, the
remaining guided power is
\begin{equation}
    P(x,y)
    =
    P_0(y)-\bar n^2 S(x,y).
    \label{eq:depletion}
\end{equation}
Since the scattered-field amplitude scales as
$\Delta n_g\sqrt{P}$, the reduction in pump amplitude is compensated by
increasing the modulation according to
\begin{equation}
    \Delta n_g(x,y)
    =
    \frac{\bar n\,g(x,y)}
         {\sqrt{P(x,y)}} .
    \label{eq:dng_final}
\end{equation}
The input distribution $P_0(y)$ sets how power is launched across the
waveguide. We consider two cases. A \emph{plane-wave} pump uses a uniform
$P_0(y)$, illuminating the full width equally; power in regions where the
grating is weak is never scattered and remains as residual guided light.
An \emph{optimised} pump instead launches an input profile proportional
to the total scattering demand along each line of constant $y$,
$P_0(y) \propto S(L_\mathrm{grat},y)$, so that the pump reaches the target depletion at the grating exit and no power is launched in areas with negligible scattering demand. This compensation is derived for the TE pump only. A subsequently launched TM pump experiences a different local scattering rate through the same grating and is therefore not compensated by the same modulation profile.

In the following and unless stated otherwise, simulations are performed at $\lambda=780\,\mathrm{nm}$ with cladding index $n_0=1.45$,
$n_\mathrm{eff}^\mathrm{TE}=1.45607$ and $n_\mathrm{eff}^\mathrm{TM}=1.45587$, corresponding to a birefringence $\Delta n_\mathrm{eff}=2\times10^{-4}$. The guided mode is modelled with a
$2\,\mu\mathrm{m}$ waist, and a maximum $10\times10\,\mathrm{mm}$ grating size is used for the long-distance high-depletion calculations to stay below the current UV-writing fabrication $\Delta n_g$ limit of $5\times 10^{-3}$.

\subsection{Effect of Pump Type and Depletion on Grating Profiles}
\label{sec:profiles}

\begin{figure}[ht]
    \centering
    \includegraphics[width=\linewidth]{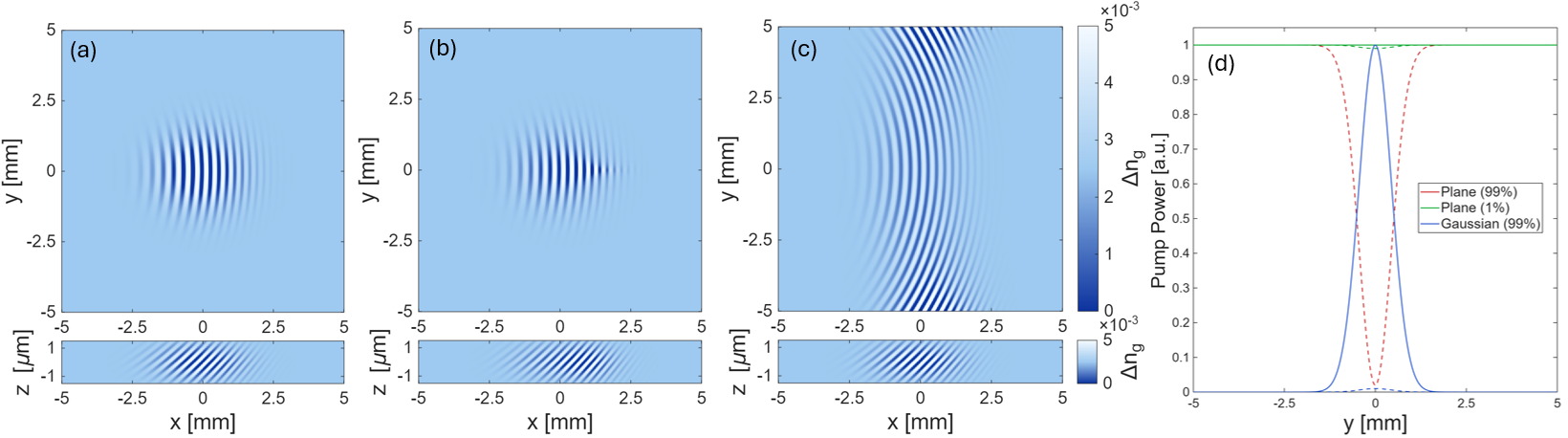}

    \caption{Grating modulation profiles (a--c) and transverse pump
    profiles (d), all targeting $w_\mathrm{tar} = 2\,\mu\mathrm{m}$ at
    $z_\mathrm{foc} = 30\,\mathrm{mm}$, target direction $(0, 0, 1)$, $\lambda=780\,\mathrm{nm}$. (a) Plane-wave pump, 1\% depletion.
    (b) Plane-wave pump, 99\% depletion. (c) Optimised
    Gaussian pump, 99\% depletion: more uniform envelope with efficient
    power extraction. (d) Solid lines: input pump profiles $P_0(y)$; 
    dashed lines: depleted profiles at the grating exit. Gratings scaled for visual clarity in (a)-(c).}
    \label{fig:grating_profiles}
\end{figure}

Before turning to polarisation, we illustrate how the depletion-compensated modulation profile $\Delta n_g(x,y)$ of Eq.~\eqref{eq:dng_final} behaves for different pump types and depletion fractions. The physical Bragg period for these gratings is $\Lambda \approx 300-400\,\mathrm{nm}$, a $10\,\mathrm{mm}$ grating contains $\sim 3\times10^{4}$ oscillations. 
For visual clarity of the oscillations, the  period has been scaled by a factor of $250$. Fig.~\ref{fig:grating_profiles} shows three representative cases, all targeting a focused Gaussian of waist $w_\mathrm{tar}=2\,\mu\mathrm{m}$ at $z_\mathrm{foc}=30\,\mathrm{mm}$ in the direction $(0,0,1)$, i.e., a vertical emission direction that requires transverse focussing across the grating. 

With a plane-wave pump and 1\% depletion, Fig.~\ref{fig:grating_profiles}(a), the pump remains essentially constant along the grating (Fig.~\ref{fig:grating_profiles}(d)) and the modulation envelope follows the target amplitude. At high depletion, Fig.~\ref{fig:grating_profiles}(b), the longitudinal variation of $P(x,y)$ reshapes the envelope into a characteristic tear-drop profile: regions where the pump has weakened require a larger modulation to maintain the scattered amplitude. The depleted pump profile in Fig.~\ref{fig:grating_profiles}(d) shows that substantial guided power remains outside the grating region.

For an optimised pump, Fig.~\ref{fig:grating_profiles}(c), power is launched only into transverse positions $y$ with significant scattering demand. The input pump profile is now confined to the region defined by Eq.~\eqref{eq:dng_shape}, in Fig.~\ref{fig:grating_profiles}(d), depleting along the grating to reach near-zero at the exit. This produces a more uniform modulation envelope and efficient power extraction, Fig.~\ref{fig:grating_profiles}(c).

\subsection{Degree of Circular Polarisation and Usable Beam Area}
\label{sec:docp}

For the gratings shown above, and in subsequent results, the design is optimised for a TE pump; consequently, the device naturally delivers mainly linear polarisation to the target. Circular polarisation requires two orthogonal components, necessitating the use of a second, TM pump. Because the two pumps deplete differently (Section~\ref{sec:depletion}), the balance between the two channels, and hence the polarisation, varies across the output beam. For the following, we therefore define the Degree of Circular Polarisation (DoCP)  as a function of position via the Stokes parameters~\cite{mcmaster_polarization_1954} as a measure of the polarisation state: At the observation plane the scattered field is decomposed into two fixed orthogonal linear components $E_s$ and $E_p$, the lab-frame $s$ and $p$ directions of the target beam, common to the whole observation plane, as distinct from the position-dependent grating-frame directions of Section~\ref{sec:alpha_s}. In terms of these,
\begin{equation}
    S_0 = |E_s|^2 + |E_p|^2, \qquad
    S_3 = -2\,\mathrm{Im}(E_s E_p^*),
    \label{eq:stokes}
\end{equation}
\begin{equation}
    \mathrm{DoCP} = S_3/S_0 \in [-1,1],
    \label{eq:docp}
\end{equation}
where $\mathrm{DoCP} = \pm1$ is pure $\sigma^\pm$ and $\mathrm{DoCP} = 0$
carries no circular component. Since driving a $\sigma^\pm$ transition
cleanly requires near-pure circular polarisation, we take
$|\mathrm{DoCP}| \geq 0.99$ as the threshold defining acceptable
polarisation purity.

\begin{figure}[h]
    \includegraphics[width=\linewidth]{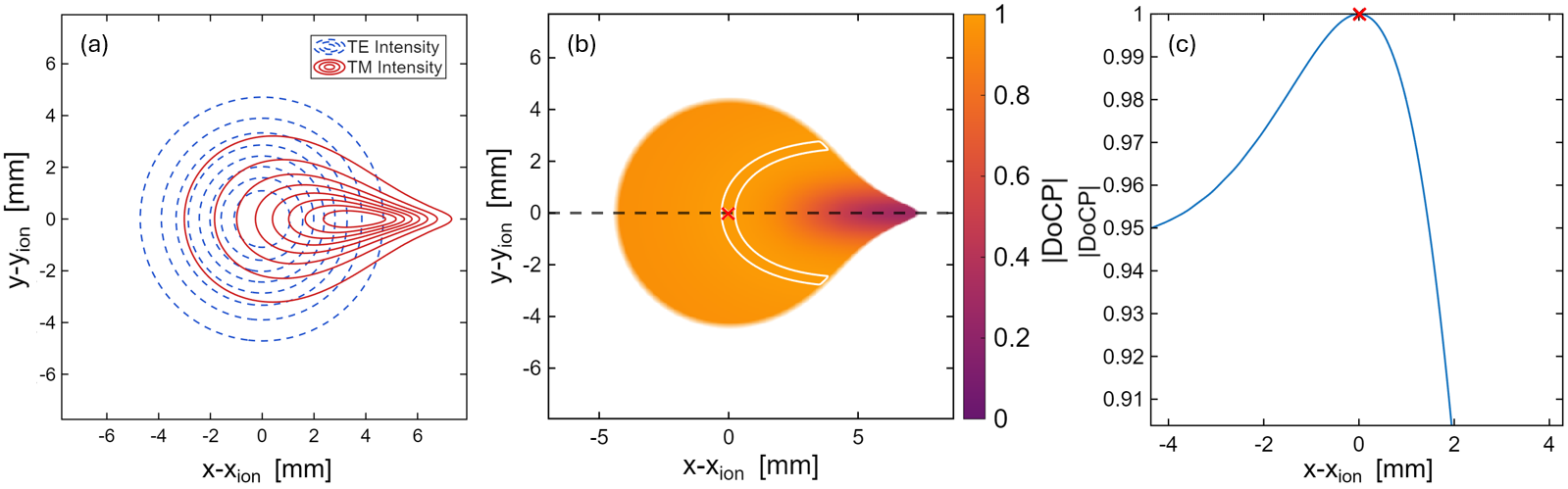}
    \caption{Intensity and DoCP at the grating plane (no propagation), $w_\mathrm{tar} = 2\,\mu\mathrm{m}$, plane-wave pump at 99\% depletion, $\lambda=780\,\mathrm{nm}$. (a) Intensity $(1/e^2)$ contours of the TE and TM fields. (b) Resulting DoCP after balancing the TE and TM channels at the beam centre (marked by $\times$) to produce circular polarisation. The white contour
    encloses the region satisfying $|\mathrm{DoCP}|\geq0.99$ and
    $I\geq I_\mathrm{peak}/e^2$. (c) Cross section of DoCP along dashed black line in (b).}
    \label{fig:docp_motivation}
\end{figure}

The distortion of the TM channel is imprinted at the grating itself: the TM
pump scatters from an envelope shaped for the TE pump, inheriting that shape
rather than the intended Gaussian. We therefore examine it at the grating
plane, where it is seen most directly and before propagation can disguise a
distortion that, being present at the source, no propagation can remove.
At $99\%$ depletion, Fig.~\ref{fig:docp_motivation}(a) and (b) illustrate the TE--TM intensity overlap and the resulting DoCP across the beam profile directly at the grating plane. Although the two channels are balanced by construction at the beam centre, yielding $|\mathrm{DoCP}| = 1$, the high-purity region collapses into a crescent that occupies a small fraction of the $1/e^2$ beam area. This source-plane distortion demonstrates that balancing the two channels at a single point gives an over-optimistic measure of robustness, motivating
a spatial measure of the usable polarisation area.

For practical devices, a single-point DoCP is insufficient to characterise the robustness of the device, since finite positioning tolerances, ion displacement and fabrication errors can shift the particle away from the design point. To capture this we define the \emph{usable circular beam area} $A$ as the region of the target plane satisfying both
\begin{equation}
    I \geq I_\mathrm{peak}/e^2
    \quad\text{and}\quad
    |\mathrm{DoCP}| \geq 0.99,
    \label{eq:thresholds}
\end{equation}
simultaneously, corresponding to the well-illuminated part of the beam that
also meets the polarisation threshold. To compare cases with a single number,
we convert $A$ to an effective radius
\begin{equation}
    R_\mathrm{eff} = \sqrt{\frac{A}{\pi}},
    \label{eq:reff}
\end{equation}
defined as the radius of a circle of equal area. For an ideal beam with
$|\mathrm{DoCP}| = 1$ across the full $1/e^2$ aperture,
$R_\mathrm{eff} = w_\mathrm{tar}$.

In the weak-depletion limit, where both polarisation channels retain the
same Gaussian profile and differ only by a transverse separation
$d_{\mathrm{sp}}$ as a consequence of waveguide birefringence, see Sec.~\ref{sec:single} B, a purely geometric reference can be obtained from the
overlap of their $1/e^2$ intensity footprints. Treating the entire overlap region as usable gives
\begin{equation}
A_\mathrm{ov}(d_{\mathrm{sp}}) =
\begin{cases}
2w_\mathrm{tar}^2
\cos^{-1}\!\left(\dfrac{d_{\mathrm{sp}}}{2w_\mathrm{tar}}\right)
-\dfrac{d_{\mathrm{sp}}}{2}
\sqrt{4w_\mathrm{tar}^2-d_{\mathrm{sp}}^2},
& d_{\mathrm{sp}} < 2w_\mathrm{tar},\\[2mm]
0, & d_{\mathrm{sp}} \geq 2w_\mathrm{tar},
\end{cases}
\label{eq:docp_walkoff}
\end{equation}

Following Eq.~\eqref{eq:reff}, the corresponding effective radius is
$R_\mathrm{eff}^{(\mathrm{walkoff})}=\sqrt{A_\mathrm{ov}/\pi}$. This provides an optimistic reference for degradation caused by geometric walk-off alone. At high depletion the TM channel becomes distorted, as shown in the grating plane in Fig.~\ref{fig:docp_motivation}, and the
degradation is no longer described by centroid separation alone. The full
simulated $R_\mathrm{eff}$ is therefore used in the following section.

\section*{Results}
% ============================================================
\section{Single-Grating Polarisation Delivery}
\label{sec:single}
% ============================================================

Section~\ref{sec:depletion} established that a single grating, designed to deplete the TE pump correctly, cannot simultaneously deplete the TM pump:
the two channels scatter differently, and the TM beam inherits the shape of the modulation envelope of the TE-optimised design. Here we ask whether any operating point of a single grating delivers usable circular polarisation to the target. 

\subsection{Fabrication limit on minimum distance}
Efficient scattering requires strong depletion, and the modulation needed
to achieve it is bounded by fabrication. From Eq.~\eqref{eq:alpha_s} the
scattering rate is local, $\alpha_s \propto \Delta n_g^2$, so the total
depletion accumulates as the pump propagates along the grating; reaching a
given depletion fraction over a longer interaction length $L_\mathrm{eff}$
requires a proportionally weaker modulation, $\Delta n_g \propto
L_\mathrm{eff}^{-1/2}$. With the index contrast capped at $\Delta n_g =
5\times10^{-3}$~\cite{holmes_planar_2011}, near-complete depletion therefore
requires a grating of order $10\,\mathrm{mm}$ long, and correspondingly a
scattered beam that occupies this length on the chip.

To generate a small target spot, the back-propagated Gaussian field must expand from its waist to a $10\,\mathrm{mm}$ footprint at the grating plane to satisfy this depletion length. Due to the divergence rate of the target Gaussian field, achieving this required expansion for a tight spot ($w_\mathrm{tar} = 2\,\mu\mathrm{m}$) forces the target plane to lie far above the chip. The inset of Fig.~\ref{fig:single_summary}(a) confirms this trade-off: for $99\%$ depletion, the required modulation meets the UV-writing cap only when the target is placed at $z \approx 30\,\mathrm{mm}$. Efficient extraction into a tightly focused spot therefore forces the target to lie tens of millimetres from the chip.

\subsection{Overlap reduction at distance due to Walk-off}

The birefringence between the TE and TM modes,
$\Delta n_\mathrm{eff} = |n_\mathrm{eff}^\mathrm{TE} - n_\mathrm{eff}^\mathrm{TM}|
= 2\times10^{-4}$, produces a transverse walk-off rate of approximately
$d_{\mathrm{sp}}/z = 0.14~\mu\mathrm{m\,mm^{-1}}$. The main panel of Fig.~\ref{fig:single_summary}(a) plots the usable circular beam area, expressed through its effective radius $R_\mathrm{eff}$, as a function of target distance. The $1\%$-depletion points (circles) closely follow the analytical walk-off prediction of Eq.~\eqref{eq:docp_walkoff}.
This agreement shows that, in the weak-depletion limit, the reduction in
usable area is dominated by geometric walk-off. The centroid separation
increases linearly with distance, reaching
$d_{\mathrm{sp}} \approx 4.2~\mu\mathrm{m}$ at $30~\mathrm{mm}$, slightly exceeding
twice the $2~\mu\mathrm{m}$ beam radius, at which point the two beams
no longer overlap and $R_\mathrm{eff}$ falls to zero. In
contrast, the $99\%$-depletion points (squares) are approximately one third of the analytical value at $10~\mathrm{mm}$ and reach zero at a shorter distance of around $20~\mathrm{mm}$, indicating an additional reduction in usable area due to depletion-induced field distortion. Fig.~\ref{fig:single_summary}(c) shows the failure directly.
At $30\,\mathrm{mm}$ the $w_\mathrm{tar} = 2\,\mu\mathrm{m}$ TE and TM intensity profiles are disjoint, and the DoCP map contains no high-purity region where the two beams would need to coincide. The TE beam is delivered accurately.

\subsection{Beam distortion for wide beams}

\begin{figure}[t]
    \centering
    % First row
    \includegraphics[width = \linewidth]{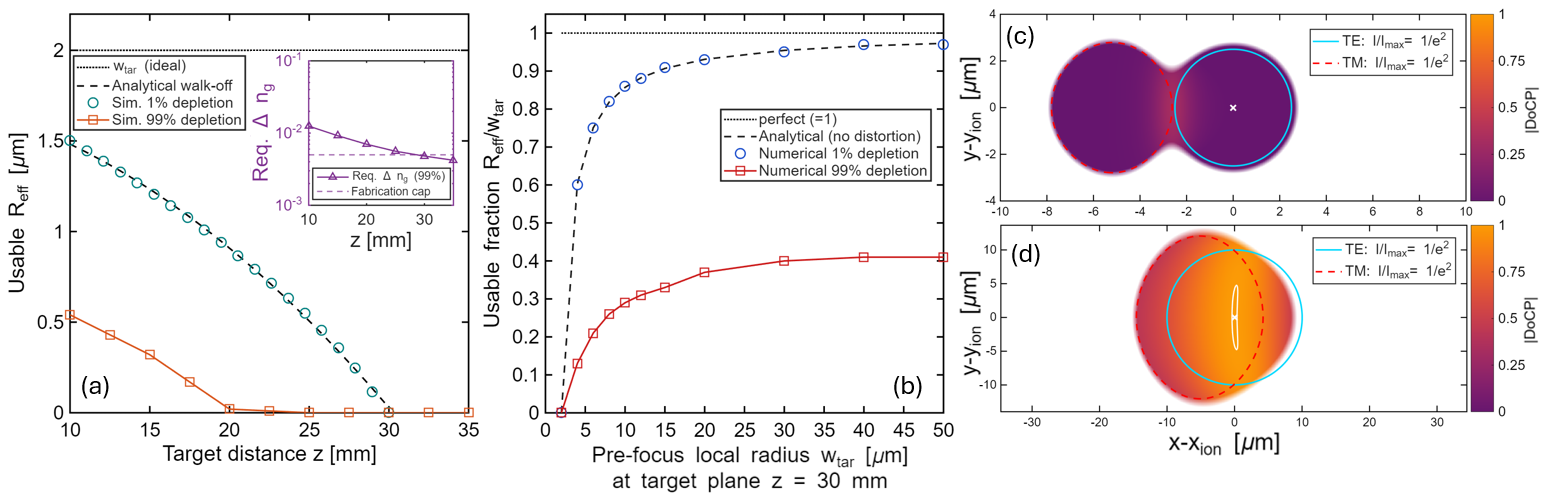}
    
    \caption{Failure of a single grating to deliver a defined polarisation state, shown for $10\,\mathrm{mm}$-long gratings with refractive index modulation capped at $\Delta n_g \leq 5\times10^{-3}$. (a) Usable circular beam area, expressed as the effective radius $R_\mathrm{eff}$, versus target distance $z$ for a Gaussian with $w_\mathrm{tar}=2\,\mu\mathrm{m}$ at $1\%$ (circles) and $99\%$ (squares) depletion. The dashed line gives the analytical geometric-overlap prediction of
    Eqs.~\eqref{eq:docp_walkoff} and~\eqref{eq:reff}. Inset: required $\Delta n_g$ for $99\%$ depletion versus distance, with the fabrication cap shown by the horizontal dashed line. (b) Usable fraction $R_\mathrm{eff}/w_\mathrm{tar}$ versus target waist $w_\mathrm{tar}$ at fixed distance $z=30\,\mathrm{mm}$, so that the walk-off separation $d_{\mathrm{sp}}$ remains fixed; the dashed line gives the same analytical geometric-overlap prediction
    at fixed $d_\mathrm{sp}$. (c) DoCP map at $z=30\,\mathrm{mm}$, $w_\mathrm{tar}=2\,\mu\mathrm{m}$ and $99\%$ depletion, with the design target marked by $\times$. The solid cyan and dashed red contours show the TE and TM $1/e^2$ scattered-intensity profiles, respectively. 
    (d) Corresponding DoCP map for $w_\mathrm{tar}=10 \,\mu\mathrm{m}$, where the broader beams are forced back into spatial overlap. The white contour encloses the region satisfying $|\mathrm{DoCP}|\geq0.99$. In all panels, the grating is designed for the TE pump, while the TM beam is left uncorrected.}

    \label{fig:single_summary}
\end{figure}

We enlarge the target waist so that the two separated beams are forced back into overlap. To maintain constant interaction length, depletion, and walk-off separation $d_\mathrm{sp}$, we keep the on-chip beam size constant but design gratings to focus on wider spot sizes at increasing distances, such that the cross section at the constant distance of $30\,\mathrm{mm}$ increases. This isolation of the target waist $w_\mathrm{tar}$ ensures that variations in $R_\mathrm{eff}$ reflect the effect of overlap alone. Figure~\ref{fig:single_summary}(b) plots the usable fraction $R_\mathrm{eff}/w_\mathrm{tar}$ against $w_\mathrm{tar}$ for $1\%$ and $99\%$ depletion levels.

At $1\%$ depletion the two beams remain approximately Gaussian, and the
usable fraction closely follows the geometric overlap prediction for their
$1/e^2$ intensity footprints, reaching $R_\mathrm{eff}/w_\mathrm{tar} \approx
0.97$ at the widest waist: a usable area of $94\%$ of the ideal. At $99\%$
depletion the usable fraction also rises but saturates at
$R_\mathrm{eff}/w_\mathrm{tar} \approx 0.40$, an area of only $16\%$ of the
ideal, and does not improve with further widening. The walk-off separation $d_{\mathrm{sp}}$ between the two channels is now reduced to a
negligible fraction of the waist, yet the TM beam is no longer Gaussian: the
tear-drop distortion imposed by the TE-optimised envelope caps the region
where the two amplitudes match. This limit is set by the mismatch in beam
\emph{shape}, which widening the waist does not remove. Fig.~\ref{fig:single_summary}(d) shows this regime: the beams now overlap, but the dashed red TM contour departs from
the solid cyan Gaussian TE profile, leaving only a thin region satisfying
$|\mathrm{DoCP}|\geq0.99$. Increasing the target waist removes the walk-off
but not the distortion, so the polarisation remains degraded.

\subsection{Conclusion on single-grating architecture}

The discussion above shows that large pump depletion, little walk-off between TE and TM polarisations, and little beam distortion cannot be achieved simultaneously: Efficient pump extraction demands high depletion, which under current fabrication constraints requires long gratings and target positions tens of millimetres away where walk-off destroys the overlap. On the other hand, widening the target beam spot increases the overlap, but then the beam distortion inherent in scattering the TE and TM pumps by a single shared grating limits the usable area to a small fraction of the target. To overcome these limitations, we introduce an orthogonal superimposed architecture in the next section.

% ============================================================
\section{Orthogonal Superimposed Grating Architecture}
\label{sec:ortho}
% ============================================================

\subsection{Architecture and Operating Principle}

The goal of the orthogonal superimposed architecture is to provide
full control of the polarisation state at the target. We have shown that the single-grating design cannot achieve this robustly because both
polarisation components must share the same modulation profile despite
their different scattering and depletion behaviour
(Section~\ref{sec:single}). We overcome this by using two superimposed
tilted gratings, each addressed by its own guided pump. As shown in
Fig.~\ref{fig:schematic_figure}(a), pump A propagates along $+x$ with a
$y$-polarised field and scatters from grating A (orange), while pump B
propagates along $+y$ with an $x$-polarised field and scatters from grating B (blue). Both gratings are designed to emit towards the same target, but their modulation profiles are optimised separately for their respective pumps. The relative amplitudes and phases of the two pumps can now be controlled independently, allowing the two scattered fields to be combined to synthesise an arbitrary polarisation state at the target.

\begin{figure}[t]
    \centering
    \includegraphics[width=\linewidth]{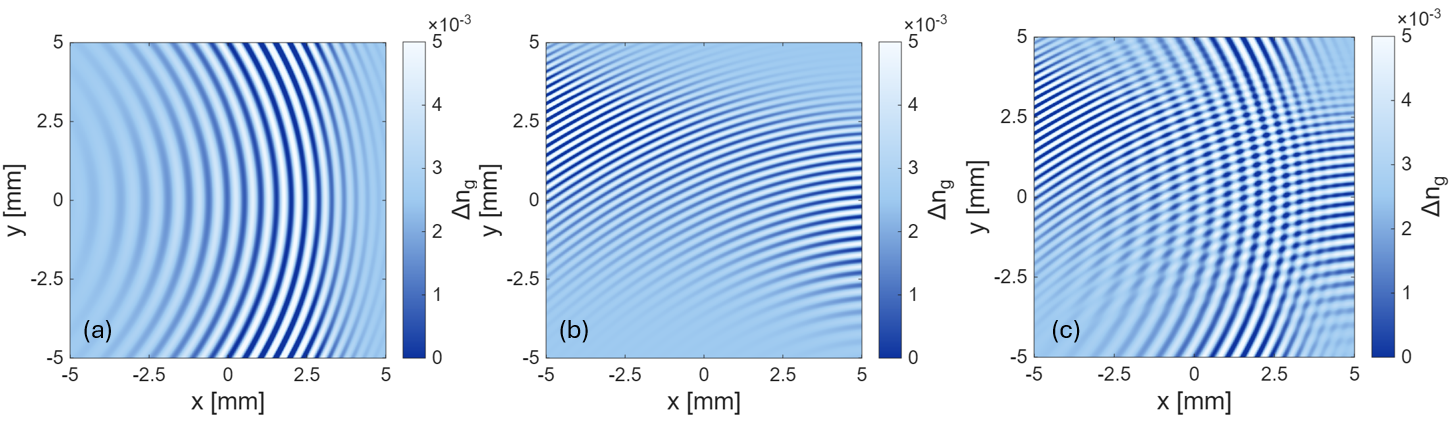}
    \caption{Grating profiles for the orthogonal architecture targeting
    $w_\mathrm{tar} = 2\,\mu\mathrm{m}$ at $z_\mathrm{foc} =
    30\,\mathrm{mm}$, direction $(0.5, 0, 1.0)$, optimised Gaussian pump at 99\% depletion, $\lambda=780\,\mathrm{nm}$.
    (a)~Grating A profile $\Delta n_{g,A}$, designed for pump A travelling
    along $+x$. (b)~Grating B profile $\Delta n_{g,B}$, designed for pump B
    travelling along $+y$. (c)~Superimposed profile
    $\Delta n_{g,A} + \Delta n_{g,B}$. Gratings scaled similarly to Fig.~\ref{fig:grating_profiles}.}
    \label{fig:cross_profiles}
\end{figure}

Fig.~\ref{fig:cross_profiles} shows the grating profiles in more detail
for a target Gaussian of waist $w_\mathrm{tar}=2\,\mu\mathrm{m}$ at
$z_\mathrm{foc}=30\,\mathrm{mm}$ in the direction $(0.5,0,1.0)$.
Panel~(a) shows grating A, designed to scatter pump A towards the target;
panel~(b) shows grating B, designed to scatter pump B to the same point;
and panel~(c) shows the resulting superimposed index-modulation profile.

A further advantage is that each pump couples to its corresponding grating through the favourable $s$-polarised scattering process, avoiding the $\cos^2(2\theta_\mathrm{inc})$ suppression of the single-grating $p$ channel. This does not mean the output of each grating is purely $s$-polarised: as established in
Section~\ref{sec:alpha_s}, any in-plane rotation $\phi \neq 0$ required for off-axis targeting projects the scattered field onto both $s$ and $p$ directions. However, the predominantly s-type scattering results in efficient beam generation and the amplitudes and
phases of the two gratings can be independently controlled.

The remaining complication is that each pump also encounters the other
grating, introducing additional cross-coupling channels alongside the two
intended scattering processes. The following sections determine where
these parasitic channels are produced and how they restrict the accessible
target-direction space.

\subsection{Parasitic Scattering Channels}
\label{sec:parasitic}

The superimposed modulation presents both spatial frequencies
$\mathbf{K}_A$ and $\mathbf{K}_B$ to each pump. Alongside the two intended scattering channels, pump A scattering on grating A (denoted as channel AA) and pump B on grating B (BB), the superimposed structure gives rise to two parasitic channels: pump A scattering from grating B (BA) and pump B scattering from grating A (AB). These are present whenever the mismatched pump--grating pair satisfies the Bragg phase-matching condition.

To identify the target directions at which parasitic scattering is strong,
we map the parasitic coupling strength across the emission direction
$(s_x, s_y) = (n_x/n_z, n_y/n_z)$, where
$\hat{\mathbf{n}}_\mathrm{target} = (n_x, n_y, n_z)$ is the desired output
direction. These maps are computed for plane-wave pumps with no depletion. Using plane waves gives a clean analytical expression for the Bragg phase-matching condition at each direction. In practice, the Gaussian pump beams used in real devices will have a finite angular bandwidth that will smooth out the sharp features visible on the maps; however, the plane-wave results correctly identify the target directions where parasitic scattering is strongest. The parasitic rate follows the same form as the $s$-polarised coefficient of Eq.~\eqref{eq:alpha_s}, evaluated for the mismatched pair. For channel AB,
\begin{equation}
    \alpha_s^{(\mathrm{AB})} \propto (\Delta n_{g,A})^2
    \exp\!\left[-\!\left(\frac{w_0}{\sin 2\theta_A}\right)^2
    \left(\beta'_B \cos^2\theta_A -
    \frac{\pi}{\Lambda_A}\right)^2\right],
    \label{eq:alpha_parasitic}
\end{equation}
where $\beta'_B = \mathbf{k}_{\mathrm{pump},B}\cdot\hat{\mathbf{n}}_A$ is
the projection of pump B's wavevector onto grating A's normal; an analogous
expression holds for BA. Large phase mismatch suppresses parasitic
scattering exponentially.

\begin{figure}[t]
  \centering
    \includegraphics[width=\linewidth]{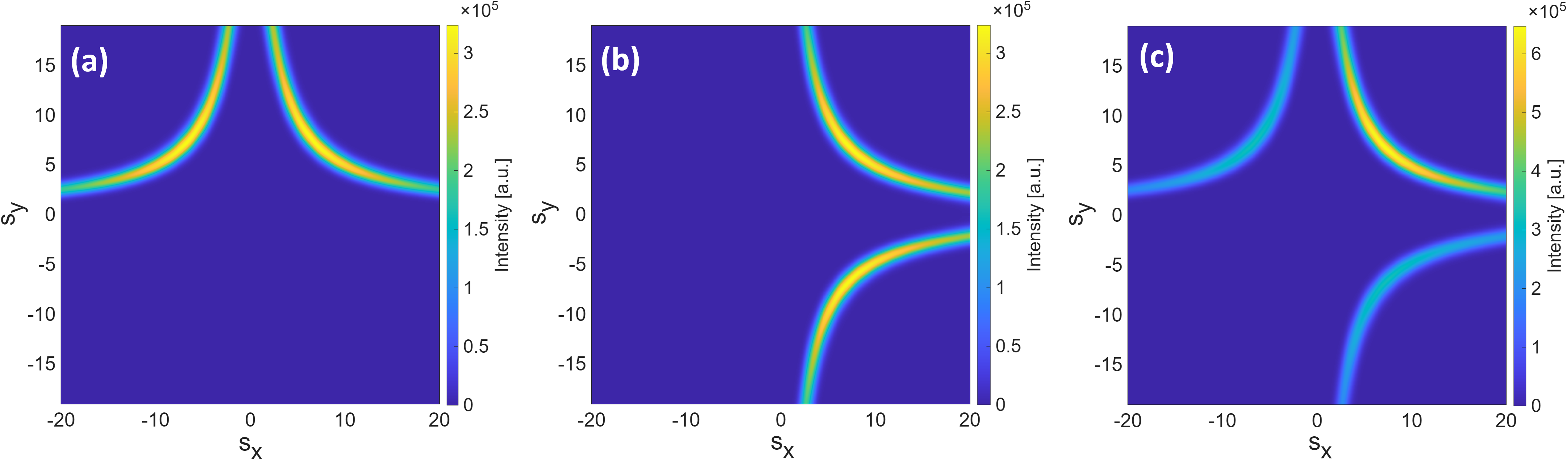}
  \caption{Parasitic scattering strength versus target direction
  $(s_x, s_y)$, computed for plane-wave pumps. (a)~BA channel (pump A on
  grating B). (b)~AB channel (pump B on grating A). (c)~Combined BA+AB
  strength. Bright arcs mark the target directions for which the mismatched
  pump--grating pair is Bragg-matched; parasitic scattering is negligible
  elsewhere.}
  \label{fig:parasitic_map}
\end{figure}

Fig.~\ref{fig:parasitic_map} shows the resulting direction maps: (a)
displays the BA channel, (b) the AB channel, and (c) their
combination. Each parasitic channel is strong along a pair of arcs where phase matching occurs. The BA arcs and AB arcs are mirror images across the diagonal, and they overlap in the upper right region of the map. Away from these arcs, i.e., for most target directions, parasitic scattering is negligible.

\subsection{Effect on the Delivered Polarisation}
\label{sec:parasitic_delivery}

The maps of Fig.~\ref{fig:parasitic_map} show target beam directions where parasitic scattering occurs, but not the propagation directions of the parasitic beam, nor how it
affects delivery of the correct polarisation state through the intended channels. These effects are coupled because AA and BA (as well as BB and AB) share their respective pumps: parasitic scattering can therefore reduce the power available to an intended channel, while its emission direction determines whether that power reaches the target.  

\begin{figure}
    \centering
    \includegraphics[width=\linewidth]{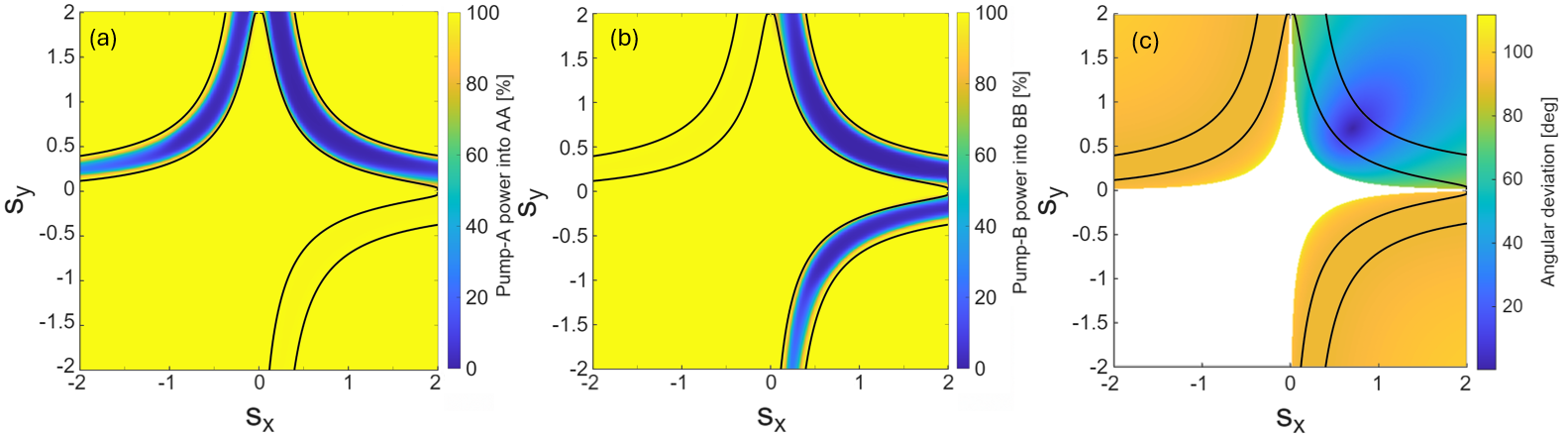}
    \caption{Delivery of the intended channels versus target direction for $w_\mathrm{tar} = 2\,\mu\mathrm{m}$, $10\,\mathrm{mm}$ gratings at $99\%$ depletion of optimised Gaussian pump, $\lambda=780\,\mathrm{nm}$, black lines mark the boundary enclosing the $|\mathrm{DoCP}| < 0.99$ region. (a)~Fraction of pump power in the intended AA channel,
    across the direction space. (b) Same as (a) but for BB channel. (c) Angular deviation between intended scattered beams and parasitic beams with same black contours as (a-b).}
    \label{fig:delivery}
\end{figure}

To quantify this, for each target direction in Fig.~\ref{fig:parasitic_map},
gratings A and B are designed to generate a circularly polarised Gaussian
beam ($w_\mathrm{tar} = 2\,\mu\mathrm{m}$) using $10\,\mathrm{mm}$ gratings at $99\%$ depletion and an optimised Gaussian pump. All four scattered channels are then propagated to the target plane, where their total and target-window powers are evaluated with shared-pump depletion included. Fig.~\ref{fig:delivery}(a) and~(b) map the fraction of each pump's power delivered to its intended channel, AA and BB respectively, across the target-direction space. Over the great majority of directions (indicated by yellow), both intended channels are fully delivered, enabling complete control over the polarisation state. Each channel falls to zero only in a small subset of directions where the corresponding pump is fully converted into its parasitic channel: pump A into BA along the arcs of panel~(a), and pump B into AB along the arcs of panel~(b). These arcs overlap in the upper-right region of direction space. On a single arc one intended channel is lost while the other is delivered in full, so only one linear component reaches the target and no circular or elliptical state can be formed whatever the relative phase; where the two arcs cross, both pumps are converted and neither intended channel is produced. Contoured on each panel (black lines) is the boundary of the region where
the full simulation gives $|\mathrm{DoCP}| < 0.99$ at the target. These
boundaries follow the parasitic arcs closely, confirming that the loss of
circularity is caused by the parasitic conversion and not by any residual
limitation of the design: at every direction outside them, the two channels are designed, propagated, and combined, and the emitted field is verified to reach the target as high-purity circular polarisation.

Whether a lost channel's converted power reaches the target depends on the
direction of the parasitic beam. Figure~\ref{fig:delivery}(c) maps the angular deviation between the two. Across almost the entire space this deviation is large: wherever a parasitic beam is produced, it propagates well away from the intended direction and would not couple to the target. Only in the small region in the upper right does the deviation fall to zero, where the parasitic beam becomes co-directional with the target. This region coincides with the arc crossing of panels~(a) and~(b), where both intended channels vanish: there, the full pump power is not only lost to the parasitic channels but is delivered, along the target direction, into the target window itself. White regions show target directions where the parasitic channel has no upward‑scattering, so the angular deviation is undefined.

The parasitic channels therefore do more than introduce stray light: on
these directions they suppress the intended channel itself, making those
targets unusable regardless of stray-light tolerance. However, the problematic
directions form narrow, predictable regions fixed by the Bragg geometry,
leaving the remainder of the accessible direction space available for
polarisation synthesis.

\section{Conclusion}

We have developed an analytical three‑dimensional scattering model, supported by numerical simulations, to evaluate how integrated tilted Bragg gratings can deliver engineered polarisation for trapped-ion and neutral-atom setups in quantum technologies. Our analysis shows that a single tilted waveguide Bragg grating is inherently limited: because both polarisation components are shaped by the same scattering structure, no configuration can generate high‑purity circular polarisation over a useful area at realistic working distances. 

To overcome this, we introduced an orthogonal superimposed‑grating architecture in 
which two independently driven gratings create an arbitrary polarisation 
state at the target. Although circular
polarisation is the state most often required, for driving $\sigma^\pm$
transitions, our design method places no restriction on the target state: any linear, circular, or elliptical polarisation can be produced by the same means. The underlying design principle is likewise
not tied to a particular beam. It applies to arbitrary spatial field
profiles, from a tightly focused spot addressing a single particle to a
broad field illuminating an extended region, and to any emission wavelength accessible to the waveguide platform. The method succeeds across almost the entire range of accessible emission directions, excluding only a narrow, predictable set fixed by the grating geometry that can easily be avoided with careful setup design.

These results demonstrate that polarisation control, traditionally achieved using macroscopic bulk optics, can be realised within a compact, chip‑integrated platform. This robust grating architecture provides a reliable route to delivering precisely engineered polarisation states for state preparation and coherent control in trapped‑ion and neutral-atom systems. We anticipate that this approach will support the development of more scalable, stable, and versatile optical control architectures for emerging quantum technologies.

\section*{Acknowledgements}
The authors would like to thank the EPSRC Centre for Doctoral Training in Quantum Technology Engineering and the QCi3 Hub at the University of Oxford.

\section*{Author contributions}
J.W.: Methodology, Simulation, Data curation, Writing – original draft, Writing – review \& editing. W.J.H.: Methodology, Analysis, Writing – review \& editing. J.G.: Methodology/Fabrication, Writing – review \& editing. P.H.: Conceptualisation, Methodology, Simulation, Supervision, Writing – review \& editing.

\section*{Funding Information}
J.W. was supported by a PhD studentship from the EPSRC Centre for Doctoral Training in Quantum Technology Engineering. W.J.H., J.G., and P.H. acknowledge financial support from the Engineering and Physical Sciences Research Council (EPSRC) via the Hub for Quantum Computing via Integrated and Interconnected Implementations (QCi3) [Grant Number EP/Z53318X/1].

\section*{Declarations}

\subsection*{Competing Interests}
The authors declare no competing interests.

\subsection{Data Availability}
Available upon reasonable request.

\bibliography{Polarisation_Paper_recent}

\end{document}